\documentclass[10pt]{article}
\usepackage{authblk}
\usepackage[utf8]{inputenc}
\usepackage[T1]{fontenc}
\usepackage{graphicx}
\usepackage{amsmath}
\usepackage{amssymb}
\usepackage{hyperref}
\usepackage[margin=30mm]{geometry}
\usepackage{parskip}
\usepackage{bm}

\newcounter{cntAlg}
\newcounter{cntStrAlg}[cntAlg]
\newlength{\algCaptUpVSkip}
\newlength{\algCaptDnVSkip}
\newlength{\algCaptHSkip}
\newlength{\algCaptAlg}
\newlength{\algStringNumWidth}
\newlength{\algStringAfterNumOffs}
\newlength{\algTabulator}
\newlength{\textAwidth}
\DeclareDocumentCommand{\algLine}{O{0} O{} m}{%
    \setlength{\textAwidth}{\textwidth}
    \noindent\parbox{\algStringNumWidth}{\refstepcounter{cntStrAlg}%
        \hfill\small\thecntStrAlg:\hspace*{\algStringAfterNumOffs}%
        \IfNoValueTF{#2}%
            {}%
            {\label{#2}}%
	}
        \IfNoValueTF{#1}%
            {}%
            {\hspace*{\dimexpr(#1\algTabulator)\relax}}%
	\parbox[t]{\dimexpr(\textAwidth-#1\algTabulator-\algStringNumWidth)\relax}{#3}%
}
\DeclareDocumentCommand{\algLineDouble}{O{0} O{} m m}{%
    \setlength{\textAwidth}{82.5mm}
    \noindent\parbox{\algStringNumWidth}{\refstepcounter{cntStrAlg}%
       \hfill\small\thecntStrAlg:\hspace*{\algStringAfterNumOffs}%
        \IfNoValueTF{#2}%
            {}%
            {\label{#2}}%
	}
        \IfNoValueTF{#1}%
            {}%
            {\hspace*{\dimexpr(#1\algTabulator)\relax}}%
	\parbox[t]{\dimexpr(\textAwidth-#1\algTabulator-\algStringNumWidth)\relax}{#3\phantom{!}}%
	\hfill
	\parbox{\algStringNumWidth}{\hfill\small\thecntStrAlg:\hspace*{\algStringAfterNumOffs}%
        \IfNoValueTF{#2}%
        {}%
        {\label{#2}}%
    }
        \IfNoValueTF{#1}%
            {}%
            {\hspace*{\dimexpr(#1\algTabulator)\relax}}%
	\parbox[t]{\dimexpr(\textAwidth-#1\algTabulator-\algStringNumWidth)\relax}{#4\phantom{!}}%
    \setlength{\textAwidth}{\textwidth}
}
\newenvironment{nmAlgorithm}[2]{\small%
\noindent\refstepcounter{cntAlg}\label{#1}\hrule\par\vspace*{\algCaptUpVSkip}%
\settowidth{\algCaptAlg}{\hspace*{\algCaptHSkip}\text{Algorithm~\thecntAlg.\ }}
\noindent\parbox[t]{\algCaptAlg}{\hfill\text{Algorithm~\thecntAlg.\ }}%
~\parbox[t]{\dimexpr(0.99\textwidth-\algCaptAlg)\relax}{\raggedright #2}\par%
\vspace*{\algCaptDnVSkip}\hrule\vspace*{\algCaptUpVSkip}\par%
}{%
\par\vspace*{\algCaptDnVSkip}\hrule\vspace*{\algCaptUpVSkip}\par%
}
\newcommand{\algINPUT}{\textbf{input\ }}
\newcommand{\algFOR}{\textbf{for\ }}
\newcommand{\algDO}{\textbf{do\ }}
\newcommand{\algENDFOR}{\textbf{end for\ }}
\newcommand{\algIF}{\textbf{if\ }}

\newcommand{\algGOTO}[1]{\textbf{goto}~\ref*{#1}}
\newcommand{\algOUTPUT}{\textbf{output\ }}
\newcommand{\algSTOP}{\textbf{stop\ }}

\title{Feynman integral reduction: balanced reconstruction of sparse rational functions and implementation on supercomputers in a co-design approach}
\author[1]{Alexander Smirnov}
\author[2]{Mao Zeng}
\affil[1]{Research Computing Center, Moscow State University, Moscow, Russia}
\affil[2]{Higgs Centre for Theoretical Physics, University of Edinburgh, Edinburgh, United Kingdom}
\date{\today}
\begin{document}
\maketitle

\begin{abstract}
Integration-by-parts (IBP) reduction is one of the essential steps in evaluating Feynman integrals. A modern approach to IBP reduction uses modular arithmetic evaluations with parameters set to numerical values at sample points, followed by reconstruction of the analytic rational coefficients. Due to the large number of sample points needed, problems at the frontier of science require an application of supercomputers. In this article, we present a rational function reconstruction method that fully takes advantage of sparsity, combining the balanced reconstruction method and the Zippel method. Additionally, to improve the efficiency of the finite-field IBP reduction runs, at each run several numerical probes are computed simultaneously, which allows to decrease the resource overhead. We describe what performance issues one encounters on the way to an efficient implementation on supercomputers, and how one should co-design the algorithm and the supercomputer infrastructure. Benchmarks are presented for IBP reductions for massless two-loop four- and five-point integrals using a development version of FIRE, as well as synthetic examples mimicking the coefficients involved in scattering amplitudes for post-Minkowskian gravitational binary dynamics.
\end{abstract}

\tableofcontents

\section{Introduction}

Feynman integral reduction is one of the essential steps in evaluating Feynman integrals, and it is based on the integration by part (IBP) relations \cite{algorithm_to_calculete_feynman_integral} that allow reducing required integrals to a small subset of so-called master integrals.

From the mathematical point of view, IBP reduction is the solving of a huge sparse system of linear equations with polynomial coefficients. This statement is a serious simplification of the process, because mathematically the number of unknowns (Feynman integrals with particular indices) and the number of relations (IBP  relations) is infinite, but it is known that the basis of the vector space of Feynman integrals of a particular family is finite-dimensional \cite{Smirnov:2010hn}. Still, when approaching a particular reduction problem, the complexity of the system might differ significantly depending on the choice of equations one is going to use --- while knowing the requested unknowns to be solved, one can choose differently the bases of what they are reduced to (the master integrals) and the intermediate relations to use. But for the most of the following paper, we are going to assume that this choice has been made, so the reduction comes down to solving a huge sparse linear system.

The conventional approach consists of solving this system directly, i.e.\ over the field of rational functions of some variables, the masses and kinematic invariants. The number of those variables might range from $1$ (only the space-time dimension) to $5\textrm{--}6$ in complicated cases. Solving such a system normally requires something like a Gaussian elimination, but on each step one needs to simplify the rational coefficients, bringing each of them to a common denominator and canceling the greatest common divisor. Avoiding such a step would lead to ``false zeros'' when a function is identically equal to zero but looks like a complex expression, or even if one could avoid false zeros, this would lead to uncontrolled growth of intermediate coefficients. There is a number of public programs preforming the conventional Feynman integral reduction \cite{Anastasiou:2004vj, Smirnov:2019qkx, Studerus:2009ye, Maierhofer:2017gsa, Klappert:2020nbg, Lee:2012cn}, some of those also using the modular approach we are going to discuss. The public programs of Refs.~\cite{Wu:2023upw, Guan:2024byi} focus on the modular approach exclusively.

The simplification of intermediate coefficients is a time-consuming step taking almost all the time in the conventional approach. A proper choice of the simplification library as well as some other tricks aim to make this step faster, but in this paper we are going to discuss another approach. One of the most promising methods of Feynman integral reduction is based on modular arithmetic and the subsequent reconstruction of analytic coefficients \cite{balancing, Monagan2004, BenOrTiw88, Zippel90, Grigoriev1994, Kaltofen88, Kaltofen90, Kaltofen2007, Kleine2005, HomoMulti2011a, Kaltofen2000, HuaGao17, Peraro:2016wsq, Peraro:2019, Klappert:2019emp, Klappert:2021, Barycentric2007, Maier:2024djk}. If one fixes specific values of variables, the calculations are to be performed over the field of rational numbers. To prevent their growth, one more step is made and one takes the projection of coefficients to a finite field modulo some large prime number. Now the coefficients can fit into machine-sized arithmetic (normally up to $2^{64}$) and solving such a system becomes immensely faster compared with the system over rational functions.

Obviously, solving a substituted system once is not enough. In order to reconstruct the analytical coefficients one needs to solve it many times for different variable values and different primes, and the number of ``sampling points'' might exceed a million. Thus the modular approach to Feynman integral reduction changes the class of the problem. The conventional approach for a complicated reduction requires a server with multiple processors, a large amount of RAM and a preferably non-interrupted run that might take a few months. The modular approach is designed for the use of supercomputers where multiple small tasks can be distributed over a set of a large number of less powerful nodes (typically with less RAM per node).

When using supercomputers, one normally encounters problems that would not appear when developing for servers, even large ones. For each supercomputer one knows its peak performance, for example, on the LINPACK benchmark. But is it possible to achieve such a performance for a particular problem? The gap between the theoretical performance and the real one might become huge, as the loss of performance comes on each step. Of course one knows that the algorithm should be optimal, but the usual starting point is to optimize in for sequential execution. When moving to a server with multiple cores, one considers parallelization of some time-consuming parts. But the prevalent approach to parallelization assumes shared memory, and its efficiency is limited by the number of cores that can exist on single server. Supercomputers are different --- the number of cores accessed might be large and exceed a thousand, but at the same time one does not get shared memory across nodes. Designing an algorithm for such a situation requires special skills and the risk to get large overhead costs for something not initially considered problematic. Furthermore, a time-consuming algorithm at the frontier of science should adapt to a particular supercomputer to efficiently utilize its specific characteristics. And ideally, the supercomputer infrastructure should also be adaptable, the supercomputer support staff should react to user problems arising in large-scale computations and adapting to them, leading to a co-design between the supercomputer and applications.

In this paper, we are going to describe a new algorithm for multivariate rational function reconstruction which fully exploits the polynomial degree bounds in individual variables and the sparsity of the rational functions. We present details of its implementation as part of the version 7 of FIRE (currently in development) and the problems (and their resolutions) encountered on supercomputers. While the code described in this paper is still private, we assume that the algorithm and the ideas presented should be of use for other scientists working with Feynman integral reduction and even outside this field.

\section{The algorithm}

\subsection{Classical Newton and Thiele methods}

Let us briefly overview the current status in reconstruction of rational functions. Not to repeat some of the formulas, we refer readers to details in the paper \cite{balancing} or Refs.\ \cite{Peraro:2016wsq, Klappert:2019emp}. The simplest case is univariate polynomial reconstruction where a simple Newton interpolation polynomial formula does the job, as there is a well-defined formula for the coefficients based on the values of the polynomial

\begin{align}
\label{eq:newton}
f_N (x) 
&= {\rm Newton}_x [f (x), N]
\\
&
\equiv 
a_1 + (x - x_1) \Big[ a_2 + (x - x_2) \big[ a_3 + (x - x_3) \left[ a_4 + \dots \right] \big] \Big]
\, . \nonumber
\end{align}

For multivariate polynomial reconstruction one can proceed variable by variable, i.e. when reconstructing from $n$ variables to $n+1$ variables one takes a needed number of reconstructed functions $f(x_1, x_2, \ldots, x_n, x_{n+1, i})$ for different values of $x_{n+1, i}$ and runs the univariate Newton reconstruction. This is not an optimal approach for sparse polynomials, and we are going to discuss it later.

When turning from polynomials to rational functions the univariate case is still simple enough for there is a replacement for the Newton formula, the so-called Thiele formula looking very similar to the previous one

\begin{align}
\label{eq:thiele}
f_T (x) 
&= {\rm Thiele}_x [f (x), T]
\\
& 
\equiv
b_0 + (x - x_1) \left[ b_1 + (x - x_2) \left[ b_2 + (x - x_3)\left[ b_4 + \dots \right]^{-1} \right]^{-1}\right]^{-1}
\, . \nonumber
\end{align}

Again, the coefficients are well-defined. The main complexity is that the Thiele formula is a continued fraction and evaluating in algebraically is much more complicated than expanding the Newton polynomial. While this still works in the univariate case, the obvious expansion to the multivariate case with an attempt to run Thiele reconstruction iteratively is inefficient due to the algebraic complexity.

Thus there exist a number of multivariate rational reconstruction methods, one of which is the balanced Newton reconstruction method proposed by one of the present authors and collaborators \cite{balancing}, which mitigates some of the inefficiencies of the homogeneous reconstruction approach suggested in Ref.~\cite{Peraro:2016wsq}.\footnote{Another alternative to the homogeneous reconstruction approach appeared recently in Ref.~\cite{Maier:2024djk} which essentially packs the exponents of variables into a single exponent with a radix numeral system.}

All the reconstruction algorithms mentioned in this paper are valid over both the field of rational numbers and finite fields modulo primes. Despite presenting some simple examples with rational numbers, all our practical implementations of the algorithms exclusively use finite fields modulo prime numbers close to $2^{64}$. After running the same reconstruction algorithm for multiple prime numbers, rational numbers will be reconstructed from the finite-field images. For this, we use Monagon's maximal quotient rational reconstruction algorithm \cite{Monagan2004} which is an improved version of Wang's algorithm \cite{wang1981p}. We will not discuss this again in the rest of the paper and will focus on the reconstruction of rational functions over finite fields. Also, we only consider generally applicable algorithms and do not exploit physics insights on the analytic structures of specific results \cite{Abreu:2018zmy, DeLaurentis:2022otd, Chawdhry:2023yyx} or relations between more than one rational functions to be reconstructed \cite{Liu:2023cgs}.

Now let us review here the main idea of the balanced Newton reconstruction method, as it will be important for the explanation of the new algorithm.

\subsection{Balanced Newton method}
\label{sec:balancedNewtonMethod}

Suppose we need to reconstruct a bivariate rational function $f(x,y) = n(x,y)/d(x,y)$ and we know its values for some $y = y_i$, being a univariate rational function in the remaining variable $x$, $n'(x, y_i)/d'(x, y_i)$ (here and later we use this notation for other functions, and not the derivative). To make the numerator and denominator of every rational function unambiguous from simultaneous scalings, we require any nontrivial polynomial GCD to have been canceled and furthermore require the ``leading monomial'' of the denominator to have a coefficient equal to one. The leading monomial is the highest-degree monomial with ties broken by e.g. lexicographic orderings in the variables involved. Take an example,
\begin{equation}
  \label{eq:fxy}
  \text{\textit{example: }} f(x, y) = \frac{n(x,y)}{d(x,y)} = \frac{xy + 2}{xy - 2x +4}, \quad n(x,y) = xy+2, \quad d(x,y) = xy - 2x + 4 \, ,
\end{equation}
where the leading monomial $xy$ of the denominator $d(x,y)$ is normalized to have unit coefficient. At $y=y_1=4$, the rational function becomes
\begin{equation}
  \label{eq:fxy_y4}
  \text{\textit{example: }} f(x, y_1=4) = \frac{4x+2} {2x+4} = \frac{2x+1}{x+2} = \frac{n'(x, y_1)} {d'(x,y_1)} \, , \quad n'(x,y_1) = 2x+1, \quad d'(x,y_1) = x+2 \, ,
\end{equation}
where after substituting $y_1=4$, we canceled a factor of $2$ between the numerator and denominator to restore the normalization convention that the leading monomial $x$ in the denominator $d'(x,y_1)$ has a unit coefficient.
One would like to run Newton reconstruction in the variable $y$ to obtain the bivariate polynomial $n(x, y)$ from the univarite polynomials $n'(x, y_i)$ known at sufficiently many values of the constant $y_i$, but this would not be possible because $n(x, y_i)$ is generally not equal to $n'(x, y_i)$, since some factors between the numerator and denominator are canceled according to the normalization convention needed to unambiguously define the numerator and denominator in the first place. We denote the cancellation with
\begin{eqnarray}
n'(x, y_i) = n(x, y_i) \cdot c(y_i)
\\
d'(x, y_i) = d(x, y_i) \cdot c(y_i)
\end{eqnarray}
where $c(y_i)$ is a rational number for each $i$. For the specific example of Eq.~\eqref{eq:fxy}, we need
\begin{equation}
  \text{\textit{example: }} c(y_i) = \frac{1}{y_i-2}
\end{equation}
to satisfy the normalization convention.
Now to be able to run Newton reconstruction one needs to derive $n(x, y_i)$ from $n'(x, y_i)$ getting rid of $c(y_i)$ or, in other words, ``balancing'' the numerator and denominator. First of all, one takes $n'(x, y_i)/n'(x_0, y_i)$ for some particular $x_0$. We have

\begin{equation}
\frac{n'(x, y_i)}{n'(x_0, y_i)} = \frac{n(x, y_i) \cdot c(y_i)}{n(x_0, y_i) \cdot c(y_i)} = \frac{n(x, y_i)}{n(x_0, y_i)} 
\end{equation}

The factor $c(y_i)$ is gone, but there is an unknown constant (for each $i$) in the denominator. Therefore we need to take one more function, substituting $x \rightarrow x_0$ for one particular value of $x$ to obtain $f(x_0, y) = n''(x_0, y)/d''(x_0, y)$. Again some factor $C = C(x_0)$ might get canceled to normalize the leading monomial of the denominator (now considered as a plynomial in y) to have a unit coefficient, so we have

\begin{eqnarray}
n''(x_0, y) = n(x_0, y) \cdot C
\\
d''(x_0, y) = d(x_0, y) \cdot C
\label{eq:nPrimeDPrime}  
\end{eqnarray}
for some constant $C$. For the example of Eq.~\eqref{eq:fxy}, we need
\begin{equation}
  \text{\textit{example: }} C = \frac 1 {x_0} \, .
\end{equation}
Now we can complete the balancing taking
\begin{equation}
\label{eq:newtonBalancing}
\frac{n'(x, y_i) \cdot n''(x_0, y_i)}{n'(x_0, y_i)} = \frac{n(x, y_i) \cdot c(y_i) \cdot n(x_0, y_i) \cdot C}{n(x_0, y_i) \cdot c(y_i)} = n(x, y_i) \cdot C \, .
\end{equation}

The coefficient $C$ on the right-hand side is a constant, since $x_0$ is considered a constant. The analogous formula works for denominators, also rescaled by the same constant $C$. Thus, knowing the numerators and denominators (according to our normalization convention) for a sequence $y_i$ and one value of $x_0$, one can balance the numerator and denominator and run two separate Newton reconstructions to obtain the rescaled numerator $n(x, y) \cdot C$ and the rescaled denominator $d(x,y) \cdot C$. Computing the ratio, we succeed in reconstructing the bivariate rational function $n(x,y)/d(x,y)$.

This works for more than two variables too, but we will skip this discussion here. 

As we demonstrated in our paper, the balanced Newton method is a useful addition to the dense reconstruction methods since it lacks some problems of homogeneous methods and requires less sampling points. But as it has been noted by reviewers and fixed in the final version of our paper, this approach is still dominated by the Zippel method designed for the sparse multivariate polynomial reconstruction. So now we are going to remind how the Zippel method works for polynomials and then we are going to present our balanced Zippel algorithm for multivariate rational function.

\subsection{Explicit Example for Balanced Newton Reconstruction}

Readers who are comfortable with the algorithm presented in Section \ref{sec:balancedNewtonMethod} can directly jump to Section \ref{sec:zippel}. Otherwise, this subsection presents a full example of reconstructing the bivariate rational function in Eq.~\eqref{eq:fxy} considered as a \emph{black box} function. For example, the black box may feed the input values of $(x,y)$ to a complicated algorithm, e.g.\ involving solving a large linear system, before returning the result. As it may be very computationally demanding to run the black box algorithm symbolically to produce the analytic form in Eq.~\eqref{eq:fxy}, we only use the black box to calculate with rational numerical values of $(x,y)$, called \emph{probes}, before reconstructing the rational function Eq.~\eqref{eq:fxy} after obtaining results for sufficiently many probes.

We arbitrary set $y_i = 3+i$ and $x_0 = 3$. First, we reconstruct $f(x, y_1=4)$ by the Thiele interpolation formula~\eqref{eq:thiele}. We evaluate $f(x, y_1=4)$ at a sequence of different values of $x$. After each evaluation, we apply Eq.~\eqref{eq:thiele} to obtain a new candidate rational function in $x$. When a new evaluation agrees with the candidate function, the procedure has \emph{converged}, and the candidate function is taken as the true function. The sequence of evaluations is show in Table \ref{tab:thieleExample}.
\begin{table}[h]
\label{tab:thiele}
\centering
 \begin{tabular}{| c | c | c | c |} 
 \hline
 $x$ & $f(x, 4)$ & Converged? & New Candidate function \\
 \hline
 3 & 7/5 & N/A & 7/5 \\ \hline
 4 & 3/2 & No & $\displaystyle x + 11/10$ \\ \hline
 5 & 11/7 & No & $\displaystyle  (2x+1)/(x+2)$ \\ \hline
 6 & 13/8 & Yes & N/A \\ \hline
 \end{tabular}
 \caption{Steps in Thiele reconstruction of $f(x, y_1=4)$ for the black box function Eq.~\eqref{eq:fxy}.}
 \label{tab:thieleExample}
\end{table}
Therefore, we have reconstructed
\begin{equation}
  f(x, y_1=4) = \frac{n'(x, y_1)} {d'(x, y_1)}, \quad n'(x, y_1) = 2x+1, \quad d'(x, y_1) = x+2 \, ,
\end{equation}
following the normalization convention that the leading monomial of the denominator has a unit coefficient. Similarly, for $y_2=5$, Thiele interpolation over the same set of $x$ values produces $$f(x, y_2) = \frac{n'(x, y_2)} {d'(x, y_2)}$$ with
\begin{equation}
  n'(x, y_2) = (5/3)x+2/3, \quad  d'(x, y_3) = x+4/3 \, .
\end{equation}
If we fix $x = x_0 = 3$ and evaluate $f(x_0, y)$ at different values of $y = 4, 5, 6, 7$, the last evaluation agrees with the candidate function produced after the preceding evaluation and gives
\begin{equation}
  f(x_0=3, y) = \frac{n''(x_0, y)} {d''(x_0, y)}, \quad n''(x_0, y) = y + 2/3, \quad d''(x_0, y) = y - 2/3 \, .
\end{equation}

According to Eq.~\eqref{eq:nPrimeDPrime}, $n''$ and $d''$ are different from $n$ and $d$, respectively, by a constant factor. Therefore, we know that $n$ and $d$ are linear polynomials in $y$.\footnote{In principle, the polynomial degree can change if there are accidental cancellations at $x=x_0$, which in practice can be avoided by choosing a large random value of $x_0$.} Then it suffices to obtain Eq.~\eqref{eq:newtonBalancing} at two different values of $y$ to complete the reconstruction, as will become clear shortly.
Having obtained the univariate rational functions, $n'(x, y_i)$ for two constant values of $y_i$ and $n''(x_0, y)$ for a constant $x_0$, we are ready to evaluate the LHS of Eq.~\eqref{eq:newtonBalancing} for different $y_i$, and this gives
\begin{equation}
  \frac{n'(x, y_i) \cdot n''(x_0, y_i)}{n'(x_0, y_i)} =
  \begin{cases}
    (4/3) x + (2/3), & y_1 = 4 \\
    (5/3) x + (2/3), & y_1 = 5 \\
  \end{cases}
  \label{eq:balancedVersusYi}
\end{equation}
Applying the Newton interpolation formula Eq.~\eqref{eq:newton} to the constant term and the $x$ coefficient in Eq.~\eqref{eq:balancedVersusYi}, we recover the polynomial dependence on y,
\begin{equation}
  \frac{n'(x, y) \cdot n''(x_0, y)}{n'(x_0, y)} = (y / 3)x + 2/3 = n(x,y) \cdot C \, .
\end{equation}
A similar procedure for the denominators yields
\begin{equation}
  \frac{d'(x, y) \cdot d''(x_0, y)}{d'(x_0, y)} = (y/3-2/3)x + 4/3 = d(x,y) \cdot C \, .
\end{equation}
Taking the ratios of the two bivariate polynomial expressions above, we cancel the unknown factor $C$ and succeed in reconstructing the analytic rational function given by the original expression Eq.~\eqref{eq:fxy}. The probe points $(x,y)$ used, not including repeatedly used ones as the black box calculation can be cached, include the following,
\begin{align*}
  & (3,4), \, (4,4), \, (5,4), \, (6,4), \\
  & (3,5), \, (4,5), \, (5,5), \, (6,5), \\
  & (3,6), \, (3,7), \\
\end{align*}
with 10 probes used in total.
\subsection{Zippel method}
\label{sec:zippel}

The Zippel algorithm for polynomials works in the following the way. Let us suppose that we have already reconstructed a polynomial $f(x_1, \ldots, x_{k-1}, c_0, \ldots)$, where $c_0$ is some constant value of $x_k$ and the remaining variables also have some fixed values. We would like to run the Newton reconstruction for $x_k$, therefore we need similar reconstructed polynomials for other values of $x_k$. One might work in a traditional manner with Newton reconstructions for previous variables, but we want to exploit the sparsity of the polynomial, i.e.\ the absence of certain monomials in the variables $(x_1, \ldots , x_{k-1})$ under a certain bound on the monomial degrees.

So suppose we take another value $c_i$ of $x_k$. We consider the already existing polynomial $f(x_1, \ldots, x_{k-1}, c_0, \ldots)$ as a skeleton, take all its non-zero monomials and assume that the set of nonzero monomials will remain the same for the yet unknown $f(x_1, \ldots, x_{k-1}, c_i, \ldots)$. Here we will leave aside the probability for this assumption to remain valid and how the values of the constants should be chosen. With this assumption we can write 

\begin{equation}
f(x_1, \ldots, x_{k-1}, c_i, \ldots) =  a_1 \cdot x_1^{p_1,1} \ldots {x_{k-1}}^{p_{k-1,1}} + \ldots +  a_t \cdot x_1^{p_1,t} \ldots {x_{k-1}}^{p_{k-1,t}}
\end{equation}
for some $t$, where $a_i$ are unknown constant coefficients and $p$ are exponents taken from the skeleton.

This is a linear system for $a_i$ and thus knowing the values of $f(x_1, \ldots, x_{k-1}, c_i, \ldots)$ for $t$ different sets of $\{x_1, \ldots, x_{k-1}\}$ (sampling points) we can solve the system. This however has a complexity $O(t^3)$ and can grow fast with the number of variables. Thus the Zippel algorithm for polynomials suggests a specific set of sampling points, i.e.\
${y_1, ... y_{k-1}}, {y_1^2, ... y_{k-1}^2}, \ldots, {y_1^t, ... y_{k-1}^t}$. In this case the systems turns into a Vandermonde system which can be solved with complexity $O(t^2)$.
The Zippel method consists of solving this system leading to the knowledge of $f(x_1, \ldots, x_{k-1}, c_i, \ldots)$ for different $i$, followed by univariate Newton reconstruction in $x_k$ to obtain $f(x_1, \ldots, x_{k-1}, x_k, \ldots)$. Proceeding iteratively, now we know the nonzero monomials in the variables $(x_1, \ldots , x_k)$, and the Zippel algorithm for polynomials can be used to reconstruct the polynomials in these $k$ variables at other numerical values of the remaining constants denoted by the dots. Note that the last variable will be only reconstructed using the Newton method, and we usually choose the last variable to be the spacetime dimension $d$, as the polynomials involved are typically dense in $d$.

\subsection{Balanced Zippel method}

The known approach in literature to apply the Zippel method to rational functions \cite{Klappert:2019emp, Klappert:2021} is based on the homogeneous approach \cite{Peraro:2016wsq}, but we are going to suggest a new method which combines the Zippel method with the balanced reconstruction method \cite{balancing}. As with the polynomial Zippel or balanced Newton methods, we are going to work variable by variable.

Suppose we need to reconstruct a rational function 

\begin{equation}
  f(x_1, \ldots, x_{k-1}, x_k, \ldots) = \frac{n(x_1, \ldots, x_{k-1}, x_k, \ldots)}{d(x_1, \ldots, x_{k-1}, x_k, \ldots)},
  \label{eq:balancedZippelToReconstruct}
\end{equation}

where the dots at the end denote some fixed values of the remaining variables. And suppose we have already reconstructed 

\begin{equation}
f(x_1, \ldots, x_{k-1}, c_0, \ldots) = \frac{n'(x_1, \ldots, x_{k-1}, c_0, \ldots)}{d'(x_1, \ldots, x_{k-1}, c_0, \ldots)}
\end{equation}
for some fixed value $c_0$ of $x_k$. Since some factor can be canceled out after the substitution we write

\begin{eqnarray}
n'(x_1, \ldots, x_{k-1}, c_0, \ldots) = n(x_1, \ldots, x_{k-1}, c_0, \ldots) \cdot C
 \\
d'(x_1, \ldots, x_{k-1}, c_0, \ldots) = d(x_1, \ldots, x_{k-1}, c_0, \ldots) \cdot C
\end{eqnarray}
for a constant C. Let us calculate the maximal number $z$ of non-zero monomials in $n'(x_1, \ldots, x_{k-1}, c_0, \ldots)$ and $d'(x_1, \ldots, x_{k-1}, c_0, \ldots)$ and assume that for other $c_i$ the set of nonzero monomials remains the same. Also, we calculate the number $t$ of sampling points needed for Thiele reconstruction of the function in the variable $x_k$ for fixed values of other variables.

Now one needs to evaluate the unknown function in $z \cdot t$ sampling points 
$f(y_1^i, \ldots, y_{k-1}^i, c_0^j)$ for some fixed $y$, with $i$ from $1$ to $z$ and $j$ from $1$ to $t$.

Then the Thiele reconstruction is performed for each $i$, resulting in a set of univarite rational functions in $x_k$,

\begin{equation}
f(y_1^i, \ldots, y_{k-1}^i, x_k, \ldots) = \frac{n_i(y_1^i, \ldots, y_{k-1}^i, x_k, \ldots)}{d_i(y_1^i, \ldots, y_{k-1}^i, x_k, \ldots)} \, .
\end{equation}

Since some factor can be cancelled, we have

\begin{eqnarray}
n_i(y_1^i, \ldots, y_{k-1}^i, x_k, \ldots) = n(y_1^i, \ldots, y_{k-1}^i, x_k, \ldots) \cdot h_i
 \\
d_i(y_1^i, \ldots, y_{k-1}^i, x_k, \ldots) = d(y_1^i, \ldots, y_{k-1}^i, x_k, \ldots) \cdot h_i
\end{eqnarray}

for some constant (but dependent on $i$) coefficient $h_i$.

Now let us create a balanced expression for the numerator for each of the needed $j$. We can write

\begin{eqnarray}
\frac{n_i(y_1^i, \ldots, y_{k-1}^i, c_0^j, \ldots) \cdot n'(y_1^i, \ldots, y_{k-1}^i, c_0, \ldots)}
{n_i(y_1^i, \ldots, y_{k-1}^i, c_0, \ldots)}  &=& \nonumber \\ 
\frac{n(y_1^i, \ldots, y_{k-1}^i, c_0^j, \ldots) \cdot h_i \cdot n(y_1^i, \ldots, y_{k-1}^i, c_0, \ldots) \cdot C}
{n(y_1^i, \ldots, y_{k-1}^i, c_0, \ldots) \cdot h_i} &=&
n(y_1^i, \ldots, y_{k-1}^i, c_0^j, \ldots) \cdot C.
\end{eqnarray}

This is our unknown numerator multiplied by a constant factor, so now we can run the Zippel polynomial reconstruction for each of the $j$, obtaining $n(x_1, \ldots, x_{k-1}, c_0^j, \ldots) \cdot C$, and a subsequent Newton reconstruction obtaining $n(x_1, \ldots, x_{k-1}, x_k, \ldots) \cdot C$. The same procedure is applied to the denominators, then the constants get canceled in the ratio and gives the desired rational function in $(x_1, \ldots, x_k)$, as in Eq.~\eqref{eq:balancedZippelToReconstruct}.

It should be also noted that space-time dimension $d$ can be placed as the last variable, and with a proper master integral choice \cite{Smirnov:2020quc, Usovitsch:2020jrk}, the denominators gets factorized into a polynomial in $d$ and a polynomial in variables other than $d$. The former polynomial can be found by reconstructing the analytic dependence on other variables at a random numerical value of $d$, while the latter polynomial can be found by reconstructing the analytic dependence on $d$ at random numerical values of other variables. With the denominator known fully, the original black box rational function can be multiplied by the denominator to give the value of the numerator for each probe calculation, and the Thiele reconstruction for the last step can be replaced with Newton reconstruction, reducing the number of needed sampling points by a factor up to two. We refer to this as the \emph{balanced Zippel method with separation}, as the $d$ dependence in separated in the denominator.

\section{Benchmark for the number of probe points}
Before discussing details of implementations, we present a benchmark for the number of probe points needed for reconstructing a particular rational function. This only depends on the algorithm itself not how it is implemented in software. We consider the bivariate rational function
\begin{equation}
  \label{eq:ratFuncBenchmark}
  \frac{(d+13)^{30} \, (y^2+9)^7 + 1} { (d-4)^{29} \, (y^2-1)^5 } \, .
\end{equation}
This function is designed to mimic the typical coefficients encountered in the IBP reduction in the 4-loop calculation for post-Minkowskian gravitational two-body dynamics in \cite{Bern:2024adl} by the present authors and collaborators. The function depends on a kinematic parameter $y$ and the spacetime dimension parameter $d$. Although actual IBP reduction coefficients are more complicated, the above test function Eq.~\eqref{eq:ratFuncBenchmark} preserves the essential features that affect the number of probes needed for reconstruction, including the polynomial degrees of the numerator and denominator in each of the variables, the factorized dependence of the denominator on $y$ and $d$, and the sparsity pattern in $y$.\footnote{In the problem of Ref.~\cite{Bern:2024adl}, all IBP coefficients have either an even parity or an odd parity under $y \rightarrow -y$. Also, the polynomial degrees in $d$ and $y$ (in either the numerator or denominator) are uncorrelated; for example, monomials with a higher power of $d$ do not have a lower maximum power of $y$, as in the test function Eq.~\eqref{eq:ratFuncBenchmark}.} Table \ref{tab:benchmark} compares the number of probes needed in the balanced Zippel algorithm presented in this paper, with or without exploiting the factorized dependence of the denominators on $d$ (separation), the number of probes needed in the balanced Newton algorithm, and the number of probes needed in the homogeneous scaling algorithm of Ref.~\cite{Peraro:2016wsq}.
\begin{table}[h]
\centering
 \begin{tabular}{| c | c |} 
 \hline
 Algorithm & Number of probes per prime \\
 \hline
 Balanced Zippel with Separation & 306 \\ \hline
 Balanced Zippel & 509 \\ \hline
 Balanced Newton & 957 \\ \hline
 Homogeneous Scaling & 1424 \\ \hline
 \end{tabular}
 \caption{Number of probes (per prime number) needed to reconstruct the bivariate rational function, Eq.~\eqref{eq:ratFuncBenchmark}, using various rational function reconstruction algorithms.}
 \label{tab:benchmark}
\end{table}
The number of probes per prime number is presented. The number of prime numbers needed for reconstruction is an independent issue, and for this problem, we need 3 prime numbers with any of these algorithms, plus a 4th prime number and a random value of $(y,d)$ used to verify the correctness of the analytic reconstruction result. Looking Table \ref{tab:benchmark} from the last row to the first row, we can see that as sparsity information is taken advantage of, first by switching on the balanced algorithm, then by switching on the Zippel algorithm, and finally by using the factorized nature of denominators, the number of probe points keeps decreasing, eventually reaching less than one quarter of the number required by the crudest algorithm.

\section{Sequential implementation}

One might invent an optimal algorithm, but it means nothing without a proper implementation. Before proceeding to parallelization one should consider an optimized sequential implementation, since any type of parallelization makes the codebase more complex, and one should first understand which parts of the algorithm need this parallelization --- there is no reason to implement a parallel version of something that takes a small part of total time.

Withing the FIRE framework we have a possibility to run reduction for different sampling points at the same time for more than $5$ years~\cite{Smirnov:2019qkx} with MPI parallelization on a supercomputer, so currently we focus on the reconstruction algorithm that comes after the IBP runs.

While implementing a sequential version of the balanced Zippel method and other reconstruction methods, we obviously had to choose a library to work with rational functions. First of all we wrote a version using our FUEL framework~\cite{FUEL} which allowed us to use different libraries and compare their speed. Since we were aiming up to $5$--$6$ variables, the only libraries that scale well to such number of variables turned to be competitors, specifically FLINT~\cite{FLINT} and Symbolica~\cite{SYMBOLICA}. While the performance turned out to be similar, FLINT has a significant advantage being completely open-source and free for use.

Preliminary tests integrating reconstruction into the FIRE framework showed that the reconstruction takes a significant part of the total time (the other part being IBP runs), thus we decided to optimize our approach by removing the middle layer FUEL and working directly with FLINT objects. A number of optimization steps was performed to reach a satisfactory performance with properly chosen FLINT structures. Since the reconstruction step can be easily separated from the other part of the reduction, it could be profiled well with Valgrind and Google Perf.

As a result we obtained a stand-alone binary that accepts a needed number of FIRE tables and runs the reconstruction producing a table with reconstructed coefficients. We are aware of the FireFly library \cite{Klappert:2019emp, Klappert:2021} for modular and rational reconstruction, but we could not adapt it to our needs. First of all, we invented a new algorithm that would require a FireFly modification. But perhaps even more importantly, we have a different ideology --- the FireFly reconstruction itself can be the main command and it can request other program to create sampling points running reduction, possibly with MPI. But in our approach we start from reduction and derive what reconstruction to call based on reduction results. The reconstruction algorithms might be called on individual nodes, so the reconstruction cannot be the master MPI task for us.

The basic FIRE algorithm to obtain the result with the modular approach can be described the following way:

\begin{nmAlgorithm}{algFIRE}            
{Modular approach to reduction}
\algLine{\algINPUT Required reduction information, variables list $x_1,\ldots,x_n$}
\algLine{Fix initial variables values $y_1,\ldots,y_n$ and first large prime $p_1$}
\algLine{\algFOR $k=1\ldots n$ \algDO}
\algLine[1]{\algFOR $j=1\ldots \infty$ \algDO}
\algLine[2]{Run reduction for $y_1,\ldots,y_k^j,\ldots,y_n$}
\algLine[2]{Try Thiele reconstruction by $y_k$ modular $p_1$}
\algLine[2]{\algIF reconstructed \algGOTO{algFIRE:gotThiele}}
\algLine[1]{\algENDFOR}
\algLine[1][algFIRE:gotThiele]{record the number of sampling points $t_j$ needed for Thiele reconstruction by $x_j$}
\algLine{\algENDFOR}
\algLine{\algFOR $p=p_1\ldots$ \algDO}
\algLine[1]{Seed $t_1$ sampling points $y_1^j,y_2,\ldots,y_n$ for $j=1..t_1$ and run reductions}
\algLine[1]{Thiele-reconstruct by $x_1$ modular $p$ (these two steps can be skipped for $p_1$ since obtained ealier)}
\algLine[1]{\algFOR $k=2\ldots n$ \algDO}
\algLine[2]{Having a reconstructed function by $x_1\ldots x_{k-1}$ record the maximum number $z_{k-1}$ of nonzero monomials in its numerator and denominator}
\algLine[2]{Run reduction for $y_1^i,\ldots,y_{k-1},y_k^j,y_{k+1}\ldots,y_n$ for $i=1\ldots z(k-1)$ and $j=1\ldots t(k)$ modular $p$}
\algLine[2]{Run Thiele reconstruction by $x_k$ for $i=1\ldots z(k-1)$}
\algLine[2]{Run balanced Zippel reconstruction to obtain reconstructed function by $x_1\ldots x_k$}
\algLine[1]{\algENDFOR}
\algLine[1]{\algIF the reconstruction can be performed from modular fields to rational numbers \algGOTO{algFIRE:end}}
\algLine{\algENDFOR}
\algLine[0][algFIRE:end]{\algOUTPUT the reconstructed function}
\algLine{\algSTOP}
\end{nmAlgorithm}

\section{Conventional parallelization}

Each parallelization attempt should first detect which parts of the algorithm really need this parallelization (normally time-consuming parts) and which ones can be parallelized (depends on the algorithm, should be a separate study). We analyzed time-consuming parts of the algorithm \ref{algFIRE} and located two major blocks: the reduction for particular sampling points and the reconstruction algorithm. 

\subsection{Optimization of the reduction algorithm}

The reduction algorithm for particular sampling points has been described in previous papers and is generally solving a sparse linear system of equations, in this case over a finite field. As an output this algorithm saves a file to disk with ``tables'' in the FIRE format for reduction results. The table files have a prefix indicating the values of variables and the prime modulus. Thus the reconstruction can be called later as a separate process relying on tables saved to disk earlier. 

While the discussion of optimal methods to solve such systems stands outside the scope of this article, we also measured the performance of individual reduction and noticed that the time used can be split into two categories --- the arithmetic operations with coefficients and what can be considered as overhead costs organizing the process of solving. The time for these categories turned to be comparable with each other (which was not clear in advance), therefore we decided to develop a version of FIRE working not with one sampling point (variable values) but with a set of sampling points. This approach allowed us to reduce the overhead costs for they remain constant no matter how many sampling points we solve together. For example, for $16$ simultaneous sampling points, the reduction time turns out to be only around $4$ times longer.

\begin{figure}
  \centering
  \includegraphics[width=0.4\textwidth]{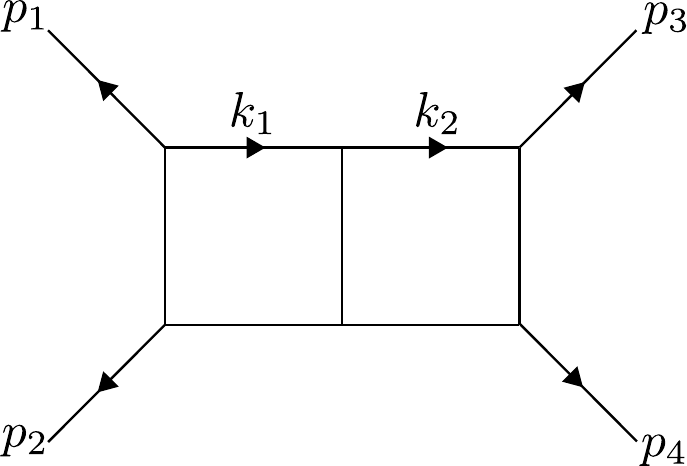}
  \caption{The double box diagram.}
  \label{fig:doublebox}
\end{figure}

We test an easy problem of the reduction of a double box integral with a rank-2 numerator (i.e.~degree 2 in the irreducible scalar products), $(k_2+p_1)^2 (k_1-p_3)^2$, as well as harder problems of the reduction of integrals with higher-rank numerators. The double box diagram is shown in Fig.~\ref{fig:doublebox}.
The kinematic variables are
\begin{equation}
  (p_1 + p_2)^2 = s, \quad (p_1+p_3)^2 = t \, .
\end{equation}
The two irreducible scalar products are chosen as
\begin{equation}
  (k_2+p_1)^2, \quad (k_1-p_3)^2 \, .
\end{equation}
The results are shown in Table \ref{tab:multipoints}.
\begin{table}[h]
\centering
 \begin{tabular}{| c | c | c | c | c |}
 \hline
 Numerator power & FIRE7p time & FIRE7p time $\times$ 64 & FIRE7mp with 64 values & Efficiency gain \\
 \hline
 2 & 4.2 & 268.8 & 110.2 & 2.43\\ \hline
 4 & 5.2 & 332.8 & 122.3 & 2.72  \\ \hline
 6 & 8.6 & 550.4 & 144.2 & 3.81 \\ \hline
 8 & 19.4 & 1241.6 & 214.0 & 5.80 \\ \hline
 \end{tabular}
 \caption{Benchmark results for IBP reductions for double box integrals in the modular approach, for integrals with a range of numerator powers. FIRE7p performs IBP reduction for one sampling point at a time, and FIRE7mp performs IBP reduction for a number (chosen to be 64) of sampling points simultaneously.}
 \label{tab:multipoints}
\end{table}

\subsection{Reconstruction optimization}

The reconstruction algorithm reads the required table files and runs the reconstruction, then saves the result to another file. There are two obvious sources for conventional parallelization that we implemented.

First of all, reading tables can be performed in parallel. Second, tables normally contain a number of coefficients that need reconstruction, so reconstructing different coefficients can be performed independently. Both steps are parallelized with a simple OpenMP approach adding only a few {\tt pragma omp} directives to the C++ code. While being simple, this method is quite efficient with an almost linear performance gain.

\section{Advancing to supercomputers}

While with the conventional parallelism it is hard to predict the complexities one is going to encounter before an implementation, with supercomputers it is almost impossible to predict them. For the modular IBP reduction part, the first thing we had to implement was an efficient MPI distribution of reduction jobs. While being more or less straightforward, this requires the ability to handle a number of situations that can happen if

\begin{itemize}
 \item one of reduction jobs crashes due to a node failure;
 \item the job is killed due do an MPI failure;
 \item the whole job is killed due to a time limit of the cluster's job scheduler;
 \item some of the table files became corrupted due to one of those failures.
\end{itemize}

When dealing with massive MPI jobs one has to keep in mind that anything strange that can happen will eventually happen. Since jobs get killed due to hardware and software problems not related to your code, they will get eventually killed at any possible place and the algorithm has to be able to recover from this place when restarted later.

The reconstruction turned out to be a lengthy step too when performed on one node while the reductions are performed on multiple nodes. To overcome this complication we had to implement an MPI version of the balanced Zippel reconstruction algorithm. Initially we tried to read the files with tables on one node and distribute coefficients that need to be reconstructed to other nodes, but this led to a dramatic performance degradation due to heavy MPI communication. 

Thus we had to switch to an approach where multiple nodes participate in reading tables, reconstructing a part of the data they read, and saving the tables. In out first implementation, all nodes read all the tables files, and it worked perfectly until approaching a million sampling points.

\subsection{Filesystem overload}

This is one of the cases where we had to adapt to a particular supercomputer or at least to supercomputers using the same filesystem type and version. The Lomonosov supercomputer~\cite{Lomonosov} uses the OSS Lustre distributed filesystem (\url{https://wiki.lustre.org/Main_Page}) which appears to be heavily loaded when the number of files in a folder becomes large enough (an order of $50$ thousand files or more). On the other hand, it works well when files are put into subfolders. When loading the filesystem heavily we even had crashes, and the communication with the support staff showed that the filesystem produced error messages like {\tt layout.c:2121:\_\_req\_capsule\_get()) @@@ Wrong buffer for field 'niobuf\_inline' (7 of 7) in format 'LDLM\_INTENT\_OPEN',...}. Partially the problem was solved when we switched to reduction in multiple sampling points at the same time, which decreased the number of tables on disk, but the problem came back for problems of a larger scale.

A possible co-design solution here would be to upgrade the filesytem to a version 2.14 of Lustre, but that would require us to stop the calculation on the whole supercomputer for a couple of weeks. Therefore we switched to adapting the code to change the way files are stored. If the total number of the files is not essential, we can split the table files into subfolders, so that depending on the maximal variable exponent in table names and the subfolder size limit, the table is placed in a corresponding subfolder. 

Also we had to change the format of tables files to allow MPI to handle them without overloading the filesystem. The reason is that MPI has a nice set of functions allowing parallel file read, where different nodes read different parts of a file. In this case the file reading internally is handled by the MPI communication with special filesystem API that does not exist on personal computers or clusters without a distributed filesystem. Thus reorganizing the files, so that different coefficients could be read by different nodes, allowed us to use those functions and decrease the filesystem load.

These two steps allowed us to overcome the filesystem limitations without upgrading it to a new version.

\subsection{Bugs in MPI and Singularity}

When exceeding a million values needed for balanced Zippel reconstruction on the Lomonosov supercomputer, we again encountered crashes happening during reconstruction. While the previous crashes happened during table loading, the new ones took place at some random moments after tables had been succesfully loaded. The stack trace always showed a problem in MPI\_Barrier like

{\tt
libc.so.6(+0x36280) [0x7ffff4f21280]
\\
\indent mca\_btl\_openib.so(+0x1546d) [0x7fffe937346d]
\\
\indent libopen-pal.so.40(opal\_progress+0x2c) [0x7ffff45875dc]
\\
\indent mca\_pml\_ob1.so(mca\_pml\_ob1\_recv+0x2a5)[0x7fffe27301c5]
\\
\indent libmpi.so.40(ompi\_coll\_base\_barrier\_intra\_tree+0x17e)[0x7ffff5ab9f0e]
\\
\indent libmpi.so.40(MPI\_Barrier+0xa8)[0x7ffff5a6f278]
}

with following trace inside FIRE. A discussion with the support staff led to a conclusion that there seemed to be a rare bug in the version of libc installed on the supercomputer, but it could be updated without stopping the whole supercomputer for a long period.

The solution we came together to was to install the Singularity modules on the supercomputer, enabling the use of a virtualization container layer to host FIRE builds inside the container. Thus FIRE could use a newer version of libc. The modules were installed and it allowed us to proceed to larger numbers of sample points for Zippel reconstruction.

\subsection{Limits and benchmarks}

The Zippel reconstruction still remains one of the most problematic points for the whole process. We already had to optimize it in a way that each coefficient could be reconstructed on a separate node, and each node used OpenMP parallelization to use all cores to handle this reconstruction in parallel. This way we could run a reconstruction job requesting around 1.7 million points for the final Zippel call. This took around 16 hours using 210 cores (15 nodes with 14 cores per each). However the problem with the Zippel algorithm is that it is quadratic in complexity by the number of points, and the reconstruction job currently needs around 5 million points and does not fit in the time limit for a job on the Lomonosov supercomputer.

\begin{figure}
  \centering
  \includegraphics[width=0.36\textwidth]{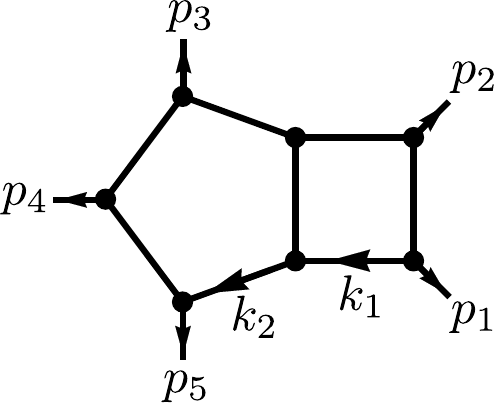}
  \caption{The penta-box diagram.}
  \label{fig:pentabox}
\end{figure}

While we cannot reveal the details of this calculation, we are going to provide some supercomputer benchmarks on an already well-known (but still complex) reduction problem (see, e.g., Ref.~\cite{Abreu:2018zmy}) of a penta-box integral with a rank-5 numerator. The penta-box diagram is shown in Fig.~\ref{fig:pentabox}.
There are five external kinematic variables, given as
\begin{equation}
  (p_1 + p_2)^2 = s_{12}, \quad (p_2 + p_3)^2 = s_{23}, \quad (p_3 + p_4)^2 = s_{34}, \quad (p_4 + p_5)^2 = s_{45}, \quad (p_5 + p_1)^2 = s_{51} \, .
\end{equation}
The irreducible scalar products are chosen to be
\begin{equation}
  (k_1 + p_3)^2, \quad (k_1 + p_5)^2, \quad (k_2 + p_1)^2 \, ,
\end{equation}
and the actual numerator of the reduced integral is
\begin{equation}
  [(k_1 + p_5)^2]^3 [(k_2 + p_1)^2]^2 \, .
\end{equation}

The finite-field version of FIRE reduces the above integral to $62$ master integrals (using only IBP identities and ignoring symmetry identities). For the purpose of the demonstration we split the master integrals into levels (the number of positive indices, i.e. uncanceled propagators) resulting in levels from $3$ to $8$. When targeting master integrals at a particular level, all other master integrals are set to zero in the IBP reduction process. While this reduction could be handled as a whole, splitting into levels is a good demonstration of how the complexity grows as one tries to obtain coefficients of master integrals at lower levels, i.e. with fewer propagators.. We ran the reduction on $210$ cores split into $15$ nodes each with $14$ cores. We used the test queue for fast job submissions having a $15$ minutes time limit. Some of the levels had to be restarted multiple times after exceeding the time limit. (The algorithm is able to restart from the point the previous run was interrupted.) We used $64$ points (in kinematic / dimension values and the modulus) per FIRE6mp run. The results are shown in Table \ref{tab:supercomputer}.
\begin{table}[h]
\centering
 \begin{tabular}{| c | c | c | c | c | c |}
 \hline
 Level & Time & Primes & Runs & Zippel & Zippel load/reconstruction time \\
 \hline
8    &      19  &   1    &    94    &  324     &   0.97 / 0.01     \\ \hline
7    &      29  &   1    &    395   &  1724    &   1.29 / 0.03   \\ \hline
6    &     107  &   1    &    1348  &   5447   &   1.79 / 0.91  \\ \hline
5    &    1689  &   2    &    8792  &   16896  &   3.40 / 7.60     \\ \hline
4    &    4358  &   2    &   26886  &   27443  &   3.51 / 21.49   \\ \hline
3    &   14968  &   2    &   29129  &   49158  &   8.41 / 68.70   \\ \hline
 \end{tabular}
 \caption{Benchmark results for the pentabox example using FIRE7mp which simultaneously calculates 64 numerical probes every run, running on 210 cores. The column \textit{Level} gives the number of propagators in every master integral that is not set to zero in the run. \textit{Time} gives the wallclock time in seconds. \textit{Primes} gives the number of primes needed for reconstruction. \textit{Runs} gives the number of FIRE7mp runs, each of which produces 64 numerical probes. \textit{Zippel} gives the size of the Vandermonde matrix involved in the last Zippel reconstruction step for reconstructing analytic dependence on kinematic variables at each numerical value of the spacetime dimension. The last column compares the IO time and the actual computation time during Zippel reconstruction, in seconds.}
 \label{tab:supercomputer}
\end{table}

While for these tests the Zippel time is still quite small, the quadradic growth speed is going to become a problem at larger numbers, requiring further work in future.

\section{Conclusion}

We described a new algorithm of the balanced Zippel reconstruction and how it fits in the general reduction scheme of Feynman integrals in the modular approach. We used this algorithm and its implementation as an example to demonstrate a co-design approach of an algorithm and supercomputers. It is in our plans to further improve the performance with all sources of parallelism including the vectorization of instructions. For the Zippel algorithm we also plan a GPU implementation. We plan to publish a paper describing a public release of the new FIRE 7 next year.

\section{Acknowledgment}
The work of Alexander Smirnov was supported in part by the Russian Science Foundation under the agreement No.\ 21-71-30003.
M.Z.’s work is supported in part by the U.K.\ Royal Society through Grant URF\textbackslash R1\textbackslash 20109. The research was carried out using the equipment of the shared research facilities of HPC computing resources at Lomonosov Moscow State University~cite{Lomonosov}. We are gratefull to the supercomputer team support, especially to I.D.~Fateev for being ready to upgrade and tune the supercomputer options for our needs. For the purpose of open access, the authors have applied a Creative Commons Attribution (CC BY) license to any Author Accepted Manuscript version arising from this submission.

\end{document}